\input harvmac
\input epsf

%
\let\includefigures=\iftrue
%
%
%
\newfam\black
\input rotate
\input epsf
\noblackbox
%
%
\includefigures
\message{If you do not have epsf.tex (to include figures),}
\message{change the option at the top of the tex file.}
\def\figin{\epsfcheck\figin}\def\figins{\epsfcheck\figins}
\def\epsfcheck{\ifx\epsfbox\UnDeFiNeD
\message{(NO epsf.tex, FIGURES WILL BE IGNORED)}
\gdef\figin##1{\vskip2in}\gdef\figins##1{\hskip.5in}
\else\message{(FIGURES WILL BE INCLUDED)}%
\gdef\figin##1{##1}\gdef\figins##1{##1}\fi}
\def\DefWarn#1{}
\def\N{{\cal N}}
\def\figinsert{\goodbreak\midinsert}
\def\ifig#1#2#3{\DefWarn#1\xdef#1{fig.~\the\figno}
\writedef{#1\leftbracket fig.\noexpand~\the\figno}%
\figinsert\figin{\centerline{#3}}\medskip\centerline{\vbox{\baselineskip12pt
\advance\hsize by -1truein\noindent\footnotefont{\bf
Fig.~\the\figno:} #2}}
\bigskip\endinsert\global\advance\figno by1}
\else
\def\ifig#1#2#3{\xdef#1{fig.~\the\figno}
\writedef{#1\leftbracket fig.\noexpand~\the\figno}%
\global\advance\figno by1} \fi

\def\tilde{\widetilde}

\def\yboxit#1#2{\vbox{\hrule height #1 \hbox{\vrule width #1
\vbox{#2}\vrule width #1 }\hrule height #1 }}
\def\fillbox#1{\hbox to #1{\vbox to #1{\vfil}\hfil}}
\def\ybox{{\lower 1.3pt \yboxit{0.4pt}{\fillbox{8pt}}\hskip-0.2pt}}

\def\da{{\dot\alpha}}

\def\rightarrowbox#1#2{
  \setbox1=\hbox{\kern#1{${ #2}$}\kern#1}
  \,\vbox{\offinterlineskip\hbox to\wd1{\hfil\copy1\hfil}
    \kern 3pt\hbox to\wd1{\rightarrowfill}}}

\def\tilp{\tilde\partial}

\def\Im{{\rm Im\hskip0.1em}}

\def\vev#1{\langle{#1}\rangle}

\def\tilde{\widetilde}

\def\II{\relax{I\kern-.10em I}}

\def\bar{\overline}

\def\IZ{\relax\ifmmode\mathchoice
{\hbox{\cmss Z\kern-.4em Z}}{\hbox{\cmss Z\kern-.4em Z}}
{\lower.9pt\hbox{\cmsss Z\kern-.4em Z}} {\lower1.2pt\hbox{\cmsss
Z\kern-.4em Z}}\else{\cmss Z\kern-.4em Z}\fi}
\def\IB{\relax{\rm I\kern-.18em B}}
\def\IC{{\relax\hbox{$\inbar\kern-.3em{\rm C}$}}}
\def\ID{\relax{\rm I\kern-.18em D}}
\def\IE{\relax{\rm I\kern-.18em E}}
\def\IF{\relax{\rm I\kern-.18em F}}
\def\IG{\relax\hbox{$\inbar\kern-.3em{\rm G}$}}
\def\IGa{\relax\hbox{${\rm I}\kern-.18em\Gamma$}}
\def\IH{\relax{\rm I\kern-.18em H}}
\def\II{\relax{\rm I\kern-.18em I}}
\def\IK{\relax{\rm I\kern-.18em K}}
\def\IN{\relax{\rm I\kern-.18em N}}
\def\IP{\relax{\rm I\kern-.18em P}}

%
\def\inbar{\,\vrule height1.5ex width.4pt depth0pt}

\font\cmss=cmss10 \font\cmsss=cmss10 at 7pt
\def\IR{\relax{\rm I\kern-.18em R}}

\def\lp10{l_P^{10}}
\def\lp11{l_P^{11}}
\def\R11{R_{11}}

\def\lt{\tilde\lambda}

\newbox\tmpbox\setbox\tmpbox\hbox{\abstractfont
}
 \Title{\vbox{\baselineskip12pt\hbox to\wd\tmpbox{\hss
 hep-th/0411107} }}
 {\vbox{\centerline{Coplanarity In Twistor Space Of $\N=4$ Next-To-MHV}
 \bigskip
 \centerline{ One-Loop Amplitude Coefficients}
 }}
\smallskip
\centerline{Ruth Britto, Freddy Cachazo, and Bo Feng}
\smallskip
\bigskip
\centerline{\it School of Natural Sciences, Institute for Advanced
Study, Princeton NJ 08540 USA}
\bigskip
\vskip 1cm \noindent

\input amssym.tex

Next-to-MHV one-loop amplitudes in $\N =4$ gauge theory can be
written as a linear combination of known multivalued functions,
called scalar box functions, with coefficients that are rational
functions. We consider the localization of these coefficients in
twistor space and prove that all of them are localized on a plane.
The proof is done by studying the action of differential operators
that test coplanarity  on the unitarity cuts of the
amplitudes.

\Date{November 2004}

\lref\BernZX{ Z.~Bern, L.~J.~Dixon, D.~C.~Dunbar and
D.~A.~Kosower, ``One Loop N Point Gauge Theory Amplitudes,
Unitarity And Collinear Limits,'' Nucl.\ Phys.\ B {\bf 425}, 217
(1994), hep-ph/9403226.
}

\lref\BernCG{ Z.~Bern, L.~J.~Dixon, D.~C.~Dunbar and
D.~A.~Kosower, ``Fusing Gauge Theory Tree Amplitudes into Loop
Amplitudes,'' Nucl.\ Phys.\ B {\bf 435}, 59 (1995),
hep-ph/9409265.
}

\lref\WittenNN{ E.~Witten, ``Perturbative Gauge Theory as a String
Theory in Twistor Space,'' hep-th/0312171.
}

\lref\CachazoKJ{ F.~Cachazo, P.~Svr\v cek and E.~Witten, ``MHV
Vertices and Tree Amplitudes in Gauge Theory,'' hep-th/0403047.
}

\lref\berkwitten{N. Berkovits and E. Witten,  ``Conformal
Supergravity In Twistor-String Theory,'' hep-th/0406051.}

\lref\penrose{R. Penrose, ``Twistor Algebra,'' J. Math. Phys. {\bf
8} (1967) 345.}

\lref\berends{F. A. Berends, W. T. Giele and H. Kuijf, ``On
Relations Between Multi-Gluon And Multi-Graviton Scattering,"
Phys. Lett {\bf B211} (1988) 91.}

\lref\berendsgluon{F. A. Berends, W. T. Giele and H. Kuijf,
``Exact and Approximate Expressions for Multigluon Scattering,"
Nucl. Phys. {\bf B333} (1990) 120.}

\lref\bernplusa{Z. Bern, L. Dixon and D. A. Kosower, ``New QCD
Results From String Theory,'' in {\it Strings '93}, ed. M. B.
Halpern et. al. (World-Scientific, 1995), hep-th/9311026.}

\lref\bernplusb{Z. Bern, G. Chalmers, L. J. Dixon and D. A.
Kosower, ``One Loop $N$ Gluon Amplitudes with Maximal Helicity
Violation via Collinear Limits," Phys. Rev. Lett. {\bf 72} (1994)
2134.}

\lref\bernfive{Z. Bern, L. J. Dixon and D. A. Kosower, ``One Loop
Corrections to Five Gluon Amplitudes," Phys. Rev. Lett {\bf 70}
(1993) 2677.}

\lref\bernfourqcd{Z.Bern and  D. A. Kosower, "The Computation of
Loop Amplitudes in Gauge Theories," Nucl. Phys.  {\bf B379,}
(1992) 451.}

\lref\cremmerlag{E. Cremmer and B. Julia, ``The $N=8$ Supergravity
Theory. I. The Lagrangian," Phys. Lett.  {\bf B80} (1980) 48.}

\lref\cremmerso{E. Cremmer and B. Julia, ``The $SO(8)$
Supergravity," Nucl. Phys.  {\bf B159} (1979) 141.}

\lref\dewitt{B. DeWitt, "Quantum Theory of Gravity, III:
Applications of Covariant Theory," Phys. Rev. {\bf 162} (1967)
1239.}

\lref\dunbarn{D. C. Dunbar and P. S. Norridge, "Calculation of
Graviton Scattering Amplitudes Using String Based Methods," Nucl.
Phys. B {\bf 433,} 181 (1995), hep-th/9408014.}

\lref\ellissexton{R. K. Ellis and J. C. Sexton, "QCD Radiative
corrections to parton-parton scattering," Nucl. Phys.  {\bf B269}
(1986) 445.}

\lref\gravityloops{Z. Bern, L. Dixon, M. Perelstein, and J. S.
Rozowsky, ``Multi-Leg One-Loop Gravity Amplitudes from Gauge
Theory,"  hep-th/9811140.}

\lref\kunsztqcd{Z. Kunszt, A. Singer and Z. Tr\'{o}cs\'{a}nyi,
``One-loop Helicity Amplitudes For All $2\rightarrow2$ Processes
in QCD and ${\cal N}=1$ Supersymmetric Yang-Mills Theory,'' Nucl.
Phys.  {\bf B411} (1994) 397, hep-th/9305239.}

\lref\mahlona{G. Mahlon, ``One Loop Multi-photon Helicity
Amplitudes,'' Phys. Rev.  {\bf D49} (1994) 2197, hep-th/9311213.}

\lref\mahlonb{G. Mahlon, ``Multi-gluon Helicity Amplitudes
Involving a Quark Loop,''  Phys. Rev.  {\bf D49} (1994) 4438,
hep-th/9312276.}

\lref\klt{H. Kawai, D. C. Lewellen and S.-H. H. Tye, ``A Relation
Between Tree Amplitudes of Closed and Open Strings," Nucl. Phys.
{B269} (1986) 1.}

\lref\pppmgr{Z. Bern, D. C. Dunbar and T. Shimada, ``String Based
Methods In Perturbative Gravity," Phys. Lett.  {\bf B312} (1993)
277, hep-th/9307001.}

\lref\GiombiIX{ S.~Giombi, R.~Ricci, D.~Robles-Llana and
D.~Trancanelli, ``A Note on Twistor Gravity Amplitudes,''
hep-th/0405086.
}

\lref\WuFB{ J.~B.~Wu and C.~J.~Zhu, ``MHV Vertices and Scattering
Amplitudes in Gauge Theory,'' hep-th/0406085.
}

\lref\Feynman{R.P. Feynman, Acta Phys. Pol. 24 (1963) 697, and in
{\it Magic Without Magic}, ed. J. R. Klauder (Freeman, New York,
1972), p. 355.}

\lref\Peskin{M.~E. Peskin and D.~V. Schroeder, {\it An Introduction
to Quantum Field Theory} (Addison-Wesley Pub. Co., 1995).}

\lref\parke{S. Parke and T. Taylor, ``An Amplitude For $N$ Gluon
Scattering,'' Phys. Rev. Lett. {\bf 56} (1986) 2459; F. A. Berends
and W. T. Giele, ``Recursive Calculations For Processes With $N$
Gluons,'' Nucl. Phys. {\bf B306} (1988) 759. }

\lref\BrandhuberYW{ A.~Brandhuber, B.~Spence and G.~Travaglini,
``One-Loop Gauge Theory Amplitudes In N = 4 Super Yang-Mills From
MHV Vertices,'' hep-th/0407214.
}

\lref\CachazoZB{ F.~Cachazo, P.~Svr\v cek and E.~Witten, ``Twistor
space structure of one-loop amplitudes in gauge theory,''
hep-th/0406177.
}

\lref\passarino{ L.~M. Brown and R.~P. Feynman, ``Radiative Corrections To Compton Scattering,'' Phys. Rev. 85:231
(1952); G.~Passarino and M.~Veltman, ``One Loop Corrections For E+ E- Annihilation Into Mu+ Mu- In The Weinberg
Model,'' Nucl. Phys. B160:151 (1979);
G.~'t Hooft and M.~Veltman, ``Scalar One Loop Integrals,'' Nucl. Phys. B153:365 (1979); R.~G.~
Stuart, ``Algebraic Reduction Of One Loop Feynman Diagrams To Scalar Integrals,'' Comp. Phys. Comm. 48:367 (1988); R.~G.~Stuart and A.~Gongora, ``Algebraic Reduction Of One Loop Feynman Diagrams To Scalar Integrals. 2,'' Comp. Phys. Comm. 56:337 (1990).}

\lref\neerven{ W. van Neerven and J. A. M. Vermaseren, ``Large Loop Integrals,'' Phys. Lett.
137B:241 (1984)}

\lref\melrose{ D.~B.~Melrose, ``Reduction Of Feynman Diagrams,'' Il Nuovo Cimento 40A:181 (1965); G.~J.~van Oldenborgh and J.~A.~M.~Vermaseren, ``New Algorithms For One Loop Integrals,'' Z. Phys. C46:425 (1990);
G.J. van Oldenborgh,  PhD Thesis, University of Amsterdam (1990);
A. Aeppli, PhD thesis, University of Zurich (1992).}

\lref\bernTasi{Z.~Bern, hep-ph/9304249, in {\it Proceedings of
Theoretical Advanced Study Institute in High Energy Physics (TASI
92)}, eds. J. Harvey and J. Polchinski (World Scientific, 1993). }

\lref\morgan{ Z.~Bern and A.~G.~Morgan, ``Supersymmetry relations
between contributions to one loop gauge boson amplitudes,'' Phys.\
Rev.\ D {\bf 49}, 6155 (1994), hep-ph/9312218.
}

\lref\RoiSpV{R.~Roiban, M.~Spradlin and A.~Volovich, ``A Googly
Amplitude From The B-Model In Twistor Space,'' JHEP {\bf 0404},
012 (2004) hep-th/0402016; R.~Roiban and A.~Volovich, ``All Googly
Amplitudes From The $B$-Model In Twistor Space,'' hep-th/0402121;
R.~Roiban, M.~Spradlin and A.~Volovich, ``On The Tree-Level
S-Matrix Of Yang-Mills Theory,'' Phys.\ Rev.\ D {\bf 70}, 026009
(2004) hep-th/0403190,
S.~Gukov, L.~Motl and A.~Neitzke,
``Equivalence of twistor prescriptions for super Yang-Mills,''
arXiv:hep-th/0404085,
I.~Bena, Z.~Bern and D.~A.~Kosower,
``Twistor-space recursive formulation of gauge theory amplitudes,''
arXiv:hep-th/0406133.
}

\lref\CachazoBY{ F.~Cachazo, P.~Svr\v cek and E.~Witten, ``Gauge
Theory Amplitudes In Twistor Space And Holomorphic Anomaly,''
hep-th/0409245.
}

\lref\DixonWI{ L.~J.~Dixon, ``Calculating Scattering Amplitudes
Efficiently,'' hep-ph/9601359.
}

\lref\BernMQ{ Z.~Bern, L.~J.~Dixon and D.~A.~Kosower, ``One Loop
Corrections To Five Gluon Amplitudes,'' Phys.\ Rev.\ Lett.\  {\bf
70}, 2677 (1993), hep-ph/9302280.
}

\lref\berends{F.~A.~Berends, R.~Kleiss, P.~De Causmaecker, R.~Gastmans and T.~T.~Wu, ``Single Bremsstrahlung Processes In Gauge Theories,'' Phys. Lett. {\bf B103} (1981) 124; P.~De
Causmaeker, R.~Gastmans, W.~Troost and T.~T.~Wu, ``Multiple Bremsstrahlung In Gauge Theories At High-Energies. 1. General
Formalism For Quantum Electrodynamics,'' Nucl. Phys. {\bf
B206} (1982) 53; R.~Kleiss and W.~J.~Stirling, ``Spinor Techniques For Calculating P Anti-P $\to$ W+- / Z0 + Jets,'' Nucl. Phys. {\bf
B262} (1985) 235; R.~Gastmans and T.~T. Wu, {\it The Ubiquitous
Photon: Heliclity Method For QED And QCD} Clarendon Press, 1990.}

\lref\xu{Z. Xu, D.-H. Zhang and L. Chang, ``Helicity Amplitudes For Multiple
Bremsstrahlung In Massless Nonabelian Theories,''
 Nucl. Phys. {\bf B291}
(1987) 392.}

\lref\gunion{J.~F. Gunion and Z. Kunszt, ``Improved Analytic Techniques For Tree Graph Calculations And The G G Q
Anti-Q Lepton Anti-Lepton Subprocess,''
Phys. Lett. {\bf 161B}
(1985) 333.}

\lref\GeorgiouBY{ G.~Georgiou, E.~W.~N.~Glover and V.~V.~Khoze,
``Non-MHV Tree Amplitudes In Gauge Theory,'' JHEP {\bf 0407}, 048
(2004), hep-th/0407027.
}

\lref\WuJX{ J.~B.~Wu and C.~J.~Zhu, ``MHV Vertices And Fermionic
Scattering Amplitudes In Gauge Theory With Quarks And Gluinos,''
hep-th/0406146.
}

\lref\WuFB{ J.~B.~Wu and C.~J.~Zhu, ``MHV Vertices And Scattering
Amplitudes In Gauge Theory,'' JHEP {\bf 0407}, 032 (2004),
hep-th/0406085.
}

\lref\GeorgiouWU{ G.~Georgiou and V.~V.~Khoze, ``Tree Amplitudes
In Gauge Theory As Scalar MHV Diagrams,'' JHEP {\bf 0405}, 070
(2004), hep-th/0404072.
}

\lref\Nair{V. Nair, ``A Current Algebra For Some Gauge Theory
Amplitudes," Phys. Lett. {\bf B78} (1978) 464. }

\lref\BernAD{ Z.~Bern, ``String Based Perturbative Methods For
Gauge Theories,'' hep-ph/9304249.
}

\lref\BernKR{ Z.~Bern, L.~J.~Dixon and D.~A.~Kosower,
``Dimensionally Regulated Pentagon Integrals,'' Nucl.\ Phys.\ B
{\bf 412}, 751 (1994), hep-ph/9306240.
}

\lref\CachazoDR{ F.~Cachazo, ``Holomorphic Anomaly Of Unitarity
Cuts And One-Loop Gauge Theory Amplitudes,'' hep-th/0410077.
}

\lref\giel{W. T. Giele and E. W. N. Glover, ``Higher order corrections to jet cross-sections in e+ e- annihilation,'' Phys. Rev. {\bf D46}
(1992) 1980; W. T. Giele, E. W. N. Glover and D. A. Kosower, ``Higher order corrections to jet cross-sections in hadron colliders,'' Nucl.
Phys. {\bf B403} (1993) 633. }

\lref\kuni{Z. Kunszt and D. Soper, ``Calculation of jet cross-sections in hadron collisions at order alpha-s**3,''Phys. Rev. {\bf D46} (1992)
192; Z. Kunszt, A. Signer and Z. Tr\' ocs\' anyi, ``Singular terms of helicity amplitudes at one loop in QCD and the soft limit
of the cross-sections of multiparton processes,'' Nucl. Phys. {\bf
B420} (1994) 550. }

\lref\seventree{F.~A. Berends, W.~T. Giele and H. Kuijf, ``Exact And Approximate Expressions For Multi - Gluon Scattering,'' Nucl. Phys.
{\bf B333} (1990) 120.}

\lref\mangpxu{M. Mangano, S.~J. Parke and Z. Xu, ``Duality And Multi - Gluon Scattering,'' Nucl. Phys. {\bf B298}
(1988) 653.}

\lref\mangparke{M. Mangano and S.~J. Parke, ``Multiparton Amplitudes In Gauge Theories,'' Phys. Rep. {\bf 200}
(1991) 301.}

\lref\grisaru{M. T. Grisaru, H. N. Pendleton and P. van Nieuwenhuizen, ``Supergravity And The S Matrix,'' Phys. Rev.  {\bf D15} (1977) 996; M. T. Grisaru and H. N. Pendleton, ``Some Properties Of Scattering Amplitudes In Supersymmetric Theories,'' Nucl. Phys. {\bf B124} (1977) 81.}

\lref\Bena{I. Bena, Z. Bern, D. A. Kosower and R. Roiban, ``Loops in Twistor Space,'' hep-th/0410054.}

\lref\BernKY{
Z.~Bern, V.~Del Duca, L.~J.~Dixon and D.~A.~Kosower,
``All Non-Maximally-Helicity-Violating One-Loop Seven-Gluon Amplitudes In N =
4 Super-Yang-Mills Theory,''
arXiv:hep-th/0410224.
}

\lref\BrittoNJ{
R.~Britto, F.~Cachazo and B.~Feng,
``Computing one-loop amplitudes from the holomorphic anomaly of unitarity
cuts,''
arXiv:hep-th/0410179.
}

\lref\BidderTX{
S.~J.~Bidder, N.~E.~J.~Bjerrum-Bohr, L.~J.~Dixon and D.~C.~Dunbar,
``N = 1 supersymmetric one-loop amplitudes and the holomorphic anomaly of
unitarity cuts,''
arXiv:hep-th/0410296.
}

\newsec{Introduction}

Perturbative amplitudes in $\N=4$ gauge theory possess many
remarkable properties. One of them is that when transformed into
twistor space \penrose, the amplitudes are localized on simple
algebraic sets \refs{\WittenNN,\RoiSpV,\CachazoKJ,\CachazoZB,\CachazoBY}.

At tree level the algebraic sets can be thought of as unions of
lines, or $\Bbb{CP}^1$'s, linearly embedded in $\Bbb{CP}^3$ \CachazoKJ. In
\WittenNN, differential operators were introduced in order to
study the support of the amplitudes directly in momentum space,
without having to compute their twistor space transform. It turns
out that at tree level, a straightforward application of
these operators probes the structure of the
amplitudes only for generic values of momenta, i.e., away from
collinear or multi-particle singularities. 
For generic values of momenta, the
twistor space picture simplifies; lines intersect in order to form
connected quivers.

In the particular case of next-to-MHV amplitudes where three
gluons have negative helicity and all other gluons have positive helicity,
there are only two lines. Therefore, if the two lines intersect, we
can say that the amplitude is localized on a plane.

At one-loop, the original proposal of \WittenNN\ suggests that the
localization on simple algebraic sets should hold. In \CachazoZB,
this structure was studied using the operators of \WittenNN, and
the result did not seem to agree with the original picture.
Motivated by the work of \BrandhuberYW, this issue was
reconsidered in \CachazoBY, where an anomaly in the action of the
operator was found to be responsible for the apparent
disagreement. Once this anomaly is taken into account, the original
picture is recovered \refs{\CachazoBY,\Bena,\CachazoDR,\BidderTX}.

One more important property of $\N=4$ amplitudes at one-loop is
that they can be written as a sum over scalar box functions with
rational functions as coefficients \refs{\passarino,\neerven,\melrose,\bernTasi,\BernKR,\morgan,\BernZX}.  For next-to-MHV amplitudes,
these coefficients can be efficiently calculated by using the
holomorphic anomaly of unitarity cuts \refs{\CachazoDR,\BrittoNJ}.
As an application of the method, the seven gluon amplitude with
helicities $(-,-,-,+,+,+,+)$ was computed. This result was
independently obtained by the direct unitarity cut method in
\BernKY, along with the results for all other helicity
configurations.

The method used in \refs{\CachazoDR,\BrittoNJ} to compute the
coefficients has as a byproduct that the coefficients are
localized on configurations in twistor space where some gluons lie
on lines.

In \BernKY, the twistor space localization of the coefficients
was considered in detail. It was found that all coefficients of
seven-gluon next-to-MHV amplitudes are localized on a plane. In
addition, the coefficient of a certain class of three-mass box
function was obtained to all multiplicities in next-to-MHV 
amplitudes with three adjacent
minuses. For $n\leq 10$, these coefficients were found
to be localized on a plane by numerical methods.  
Finally, the authors of \BernKY\ presented an outline of a proof
that the coefficients of all next-to-MHV amplitudes are localized on a plane.

It is the aim of this paper to prove this statement.  
Our proof is significantly different from the argument of \BernKY, 
but both are based on extending the arguments of \CachazoDR, 
which shows that the
coefficients are necessarily annihilated by some collinear
operators, to include coplanar operators.

This paper is organized as follows: In section 2, we
 show that proving that the coefficients 
are annihilated by a certain differential operator
is equivalent to proving that the  operator produces a
rational function when acting on certain unitarity cuts of the
amplitude.   This equivalence applies to all one-loop amplitudes.
In section 3, we prove the latter statement for the coplanar operator acting on next-to-MHV amplitudes. Finally,
in the appendix we prove that at tree-level, next-to-MHV amplitudes
of gluons with at most two fermions or scalars are coplanar. This
fact is used in the proof presented in section 3.

\newsec{Preliminaries}

Any leading-color $n$-gluon ${\cal N}=4$ amplitudes at one-loop
can be written as a linear combination of scalar box functions as
follows \refs{\BernKR,\BernZX}:
\eqn\gene{\eqalign{  A_{n;1}^{\rm 1-loop} = \sum_{i=1}^n & \left(
b_i F^{1m}_{n:i}+ \sum_r c_{r,i} F^{2m~e}_{n:r;i}+ \sum_r d_{r,i}
F^{2m~h}_{n:r;i}+ \right. \cr &~~~~ \left. \sum_{r,r'} g_{ r,r',i}
F^{3m}_{n:r:r';i} +\sum_{r,r',r''} f_{ r,r',r'',i}
F^{4m}_{n:r:r':r'';i} \right).}}
The explicit form of these functions is not relevant in our
discussion.\foot{See appendix A of \CachazoDR\ for a definition of
these functions and a discussion of their discontinuities.} 
The
coefficients are rational functions of the spinor inner products
of external gluons. Recall that the momentum of each gluon can be
written as $p_{a\dot a} = \lambda_a\tilde\lambda_{\dot a}$. The
inner products are defined as follows: $\vev{\lambda ~\lambda'} =
\epsilon_{ab}\lambda^a\lambda'^b$ and   $[\tilde\lambda
~\tilde\lambda'] = \epsilon_{{\dot a}{\dot b}}\tilde\lambda^{\dot
a}\tilde\lambda'^{\dot b}$. We follow the conventions of
\WittenNN.

We want to study the twistor space support of the coefficients.
Doing so directly from the amplitude \gene\ is not simple.
However, one of the observations in \CachazoDR\ is that from
studying the action of collinear operators on the unitarity cuts
of \gene, one finds that the coefficients are annihilated by them.
Here we want to generalize this argument starting with the
following observation.

\vskip0.1in

{\it Observation:} Let ${\cal O}$ be any $k$-th order
differential operator in the spinor variables. Let $C_{i,i+1,
\ldots , j}$ denote the unitarity cut of \gene\ in the
$(i,i+1,\ldots , j)$ channel. If ${\cal O}C_{i,i+1, \ldots , j}$
is a rational function, then ${\cal O}(c) = 0$ for all
coefficients $c$ whose scalar box functions participate in this
cut.

To prove this we make an argument similar to the one in
\CachazoDR\ for collinear operators.  Recall that the unitarity
cut can be expressed two ways.  One is by the cut integral, which
in the $(i,i+1,\ldots ,j-1,j)$-channel is given by 
\eqn\cutIn{ \eqalign{ & C_{i,i+1,\ldots ,j-1,j} = \cr &   \int
d\mu ~A^{\rm tree}((-\ell_1),i,i+1,\ldots ,j-1,j,(-\ell_2))A^{\rm
tree}(\ell_2,j+1,j+2,\ldots ,i-2,i-1,\ell_1),}}
where $d\mu$ is the Lorentz invariant phase space measure of two
light-like vectors $(\ell_1, \ell_2)$ constrained by momentum
conservation.

The other way to write this unitarity cut is as the imaginary part
of the amplitude in the regime where $(p_i+p_{i+1}+\cdots+p_j)^2 >
0$ and all other kinematical invariants are negative.  This
operation selects a subset of the scalar box functions, along with
their proper coefficients.  So, when the operator ${\cal O}$ acts
on a term of the cut, schematically we find that
\eqn\julos{{\cal O}C_{i,\ldots ,j} = \sum {\cal O}\left( c~ \Im F
\right). }
Collect the terms where
no derivatives act on $\Im F$. This simply gives
\eqn\allonc{ \sum {\cal O}(c) \Im F.}

Now we use the fact that $\Im F$ is always the logarithm of a
rational function (or for the four-mass box function, the logarithm of a function of the form $A+\sqrt{B}$, where $A$ and $B$ are rational functions).
This implies that all terms where at least one
derivative acts on the logarithm do not involve logarithms.

Now recall that ${\cal O}C_{i,i+1, \ldots , j}$ is a rational
function by hypothesis. On the other hand, the terms in \allonc\
have logarithms. As shown in \CachazoDR, there is no way the
logarithms can conspire to cancel among various box functions.
Therefore, ${\cal O}(c)=0$ for each $c$ in \allonc.

In the next section we specialize to the case of next-to-MHV amplitudes where ${\cal O}$ is
a second order differential operator that tests the coplanarity of
four gluons in twistor space.

\newsec{Proof Of Coplanarity Of Coefficients In NMHV Amplitudes}

In this section, we prove that any coplanar operator in the external
gluons annihilates all of the coefficients in \gene\ for a
next-to-MHV amplitude. The idea is to show that any coplanar
operator of four external gluons acting on a unitarity cut
produces a rational function.
It follows that the coefficients in
that cut are coplanar, by the observation proven in section 2.

A coplanar operator is of the form \WittenNN\
\eqn\hereisk{K_{ijkl}= \vev{i~j}[\tilp_k~\tilp_l]+
\vev{j~k}[\tilp_i~\tilp_l] + \vev{k~i}[\tilp_j~\tilp_l]+
\vev{k~l}[\tilp_i~\tilp_j] + \vev{i~l}[\tilp_j~\tilp_k]+
\vev{j~l}[\tilp_k~\tilp_i],}
where
\eqn\ouros{ (\tilp_i)_\da = {\del \over {\del \lt^\da_i}}.}

In the remainder of this section, we prove that $K(C_{i,i+1,\ldots
,j-1,j})$ is a rational function using the integral form \cutIn\
of the unitarity cuts.

For an arbitrary next-to-MHV helicity assignment, the integral
vanishes unless one of the two tree-level amplitudes in \cutIn\ is
MHV \BrittoNJ.  The other tree amplitude will then always be
next-to-MHV.
It was shown explicitly in Appendix B of \BrittoNJ\ that every
coefficient in \gene\ can be calculated from some cut of this form, 
where the MHV side has at least three external gluons. The
next-to-MHV side has at least four, otherwise it also becomes MHV
or ${\bar {\rm MHV}}$. (If one side of the cut has only three
external gluons, it can be given an MHV assignment, or else it
vanishes.) A special case arises when there are only six gluons,
so that both of the tree amplitudes are MHV.  This will not spoil
our argument; simply exchange the two sides where appropriate.

Let us then decompose the cut \cutIn\ according to the helicity
assignments of the cut propagators and treat each contribution
separately.  Each term takes the form
\eqn\cutpiece{\int d\mu~ A^{\rm tree}_{\rm
MHV}((-\ell_1),i,i+1,\ldots ,j-1,j,(-\ell_2))A^{\rm tree}_{\rm
NMHV}(\ell_2,j+1,j+2,\ldots ,i-2,i-1,\ell_1).}

Our proof relies on the special properties of collinear and
coplanar operators acting on tree-level MHV and next-to-MHV amplitudes.
Recall that, for generic values of external momenta, collinear and
coplanar operators annihilate MHV tree amplitudes of gluons
\WittenNN, and coplanar operators annihilate next-to-MHV
tree amplitudes also \CachazoKJ.  We review and extend the proof of
these properties to the case where $\ell_1$ and $\ell_2$ are
fermions or scalars in the appendix.

Now we are ready to show that a coplanar operator, acting on each
term \cutpiece\ of the cut, produces a rational function.  Let
$m_1,m_2,\ldots$ and $n_1,n_2,\ldots$ label external gluons on the
MHV and NMHV sides of the cut, respectively.  Indices $a,b,c,d$
will label any gluons on either side. Since the coplanar operator
is antisymmetric under index exchange, there are five cases we
must consider:
$K_{m_1m_2m_3m_4},K_{m_1m_2m_3n},K_{m_1m_2n_1n_2},K_{mn_1n_2n_3},K_{n_1n_2n_3n_4}.$

For the first two cases, $K_{m_1m_2m_3m_4}$ and $K_{m_1m_2m_3n}$,
there is a useful identity that expresses the simple geometrical
fact that any three points that are collinear are coplanar with
any fourth point. This identity can be written in the following
form: \eqn\ktof{K_{abcd}=[F_{abc}~\tilp_d]+ {1 \over \vev{a~c}}
\left(
\vev{d~a}[F_{abc}~\tilp_c]+\vev{c~d}[F_{abc}~\tilp_a]\right).} We
see that the collinear operator $F_{m_1m_2m_3}$ will appear in
every term of \ktof\ for the operators $K_{m_1m_2m_3m_4}$ and
$K_{m_1m_2m_3n}$.  But the action of this collinear operator is
familiar \CachazoBY:  it acts only on the MHV side of the cut, where it
annihilates $A^{\rm tree}_{\rm MHV}$ except for the holomorphic
anomaly, which produces a delta function localizing the integral
to give a rational function.

For the fifth case, $K_{n_1n_2n_3n_4},$ the derivatives
$\tilp_{n}$  act trivially on the measure $d\mu$, so this coplanar
operator passes through the measure and the MHV amplitude. To see
this, note that the measure can be written as
follows:
\eqn\meas{ d\mu = \delta^{(+)}(\ell_1^2)
\delta^{(+)}(\ell_2^2) \delta^{(4)}(\ell_1+\ell_2-P_L), }
where $P_L = p_i+p_{i+1}+\ldots + p_j$.  If we use conformal
invariance to set the coordinates $Z=(\lambda_1,\lambda_2,\mu_{\dot 1},\mu_{\dot 2})$ in twistor space of $n_1$ and
$n_2$ to $Z_{n_1}=(1,0,0,0)$ and $Z_{n_2}=(0,1,0,0)$, then, as described in
rigorous detail in section 3.3 of \WittenNN,
\eqn\copes{ K_{n_1n_2n_3n_4} = [\tilp_{n_3}~\tilp_{n_4}].}
Moreover, $\lambda$ and $\tilde\lambda$ of the gluons in $P_L$ and
$n_3$ and $n_4$ are all independent variables, since momentum
conservation is satisfied by making $\lt_{n_1}$ and $\lt_{n_2}$ 
depend on all other gluons.

Now we can prove that $K_{n_1n_2n_3n_4}$ produces a rational
function when acting on the cut integral $C_{i,i+1,\ldots , j}$
given in \cutIn. The proof is as follows. We have shown that
$K_{n_1n_2n_3n_4}$ when acts on $C_{i,i+1,\ldots , j}$ only
affects the NMHV amplitude $A^{\rm
tree}(\ell_2, j+1, \ldots , i-1, \ell_1)$.
The particles running in the cut propagators, i.e., $\ell_1$ and
$\ell_2$, can be gluons, fermions or scalars. In any of these cases
we prove in the appendix that for generic values of $p_{\ell_2},
p_{j+1}, \ldots , p_{i-1},p_{\ell_1}$, the operator
$K_{n_1n_2n_3n_4}$ annihilates the amplitude. Since we are
assuming that the values of $p_{j+1}, \ldots , p_{i-2}$, and
$p_{i-1}$ are generic and that the cut is in at least a three-particle channel, the only possibility of a nonzero
contribution is at isolated points in the phase space integral
over $\ell_1$ and $\ell_2$ \CachazoBY.
This implies that the integral localizes and can produce
at most a rational function, since the tree-level amplitudes in the
integrand are rational.

Finally, there are the third and fourth cases, $K_{m_1m_2n_1n_2}$
and $K_{mn_1n_2n_3}$.  We will apply the results from the other
cases to prove that these operators, too, produce rational
functions.

The terms of a coplanar operator \hereisk\ are of three types:
$f_1[\tilp_{m_1}~\tilp_{m_2}], f_2[\tilp_{m}~\tilp_{n}],$ and
$f_3[\tilp_{n_1}~\tilp_{n_2}]$.  The coefficients $f_r$ of the
differential operators  are already rational, so we consider only
the action of the operators $\tilp$.   We will be able to prove
that in fact {\it each} of these terms acts on the cut to produce
a rational function.

Now, for terms of the form $f_1[\tilp_{m_1}~\tilp_{m_2}]$ (the first
type of term mentioned above),  use 
conformal invariance to set
 $Z_{m_3}=(1,0,0,0)$ and $Z_{n}=(0,1,0,0)$, where $m_3$ is different from $m_1$ and $m_2$. 
This is possible because the MHV side has at
least three external gluons. Then we find that
\eqn\dldl{[\tilp_{m_1}~\tilp_{m_2}]=K_{m_1m_2m_3n}.} This is an
operator we have already considered.

For terms of the form $f_2[\tilp_{m_1}~\tilp_{n}]$, we can make a
similar argument, using two additional gluons $m_2,m_3$ with $Z_{m_2}=(1,0,0,0)$ and $Z_{m_3}=(0,1,0,0)$,
so that \eqn\dldl{[\tilp_{m_1}~\tilp_{n}]=K_{m_1nm_2m_3},} which
again is a coplanar operator we have already considered.

Finally, there are terms of the form $f_3[\tilp_{n_1}~\tilp_{n_2}]$.
Since the next-to-MHV amplitude must have at least four external
gluons, we may choose two others $n_3,n_4$ and set
$Z_{n_3}=(1,0,0,0)$ and $Z_{n_4}=(0,1,0,0)$,
so that \eqn\drdr{[\tilp_{n_1}~\tilp_{n_2}]=K_{n_1n_2n_3n_4},} which
again is a coplanar operator we have already considered.

\bigskip
\bigskip
\centerline{\bf Acknowledgements}

It is a pleasure to thank D. Kosower for helpful discussions.
R. B. and B. F. were supported by NSF grant PHY-0070928. F. C. was
supported in part by the Martin A. and Helen Chooljian Membership
at the Institute for Advanced Study and by DOE grant
DE-FG02-90ER40542.

\appendix{A}{Coplanarity Of Next-to-MHV Tree-Level Amplitudes}

In this appendix we review the proof that all next-to-MHV tree-level
amplitudes of gluons are localized on a plane. 
We extend the proof to include ${\cal N}=4$
 amplitudes with at most two external fermions or two
external scalars.

All these amplitudes can be computed using extensions of the MHV
diagrams of \CachazoKJ\ to the case of fermions and scalars
\refs{\GeorgiouWU,\WuFB,\WuJX,\GeorgiouBY}. The basic idea is to use MHV vertices
continued off-shell and connected by propagators. Each MHV vertex
contains at most two fermions or two scalars and can easily be
computed in terms of an MHV vertex of only external gluons by using
one of the following Ward identities:
\eqn\ward{\eqalign{ A(F^-_1,g_2^+,\ldots, g_i^-, \ldots , F_n^+) =
& {\vev{j~n}\over \vev{j~1}} A^{\rm MHV }(g_1^-,g_2^+,\ldots
,g_j^-,\ldots , g_n^+ ), \cr A(S^-_1,g_2^+,\ldots, g_i^-, \ldots ,
S_n^+) = & {\vev{j~n}^2\over \vev{j~1}^2} A^{\rm MHV
}(g_1^-,g_2^+,\ldots ,g_j^-,\ldots , g_n^+ ).  } }

From this we conclude that all NMHV amplitudes can be written as
a sum of MHV diagrams of gluons with prefactors that depend only
on the holomorphic spinors. Therefore, proving that $K$
annihilates the amplitude is equivalent to proving that $K$
annihilates any of the MHV diagrams that involve only gluons and
two MHV vertices.

In general, the twistor space localization of tree amplitudes of
gluons was considered in section 2 of \CachazoZB. Here we review
the arguments of \CachazoZB\ for the particular case of tree-level
next-to-MHV amplitudes. 

A general NMHV amplitude of the form $A^{\rm
tree}_n(1^-,2^+,\ldots ,i^-,\ldots, j^-, \ldots , n)$ can be
computed by adding up all possible MHV diagrams with one link and
two nodes \CachazoKJ.

\ifig\convi{An MHV diagram contributing to the NMHV amplitude
$A^{\rm tree}_n$.}
{\epsfxsize=0.55\hsize\epsfbox{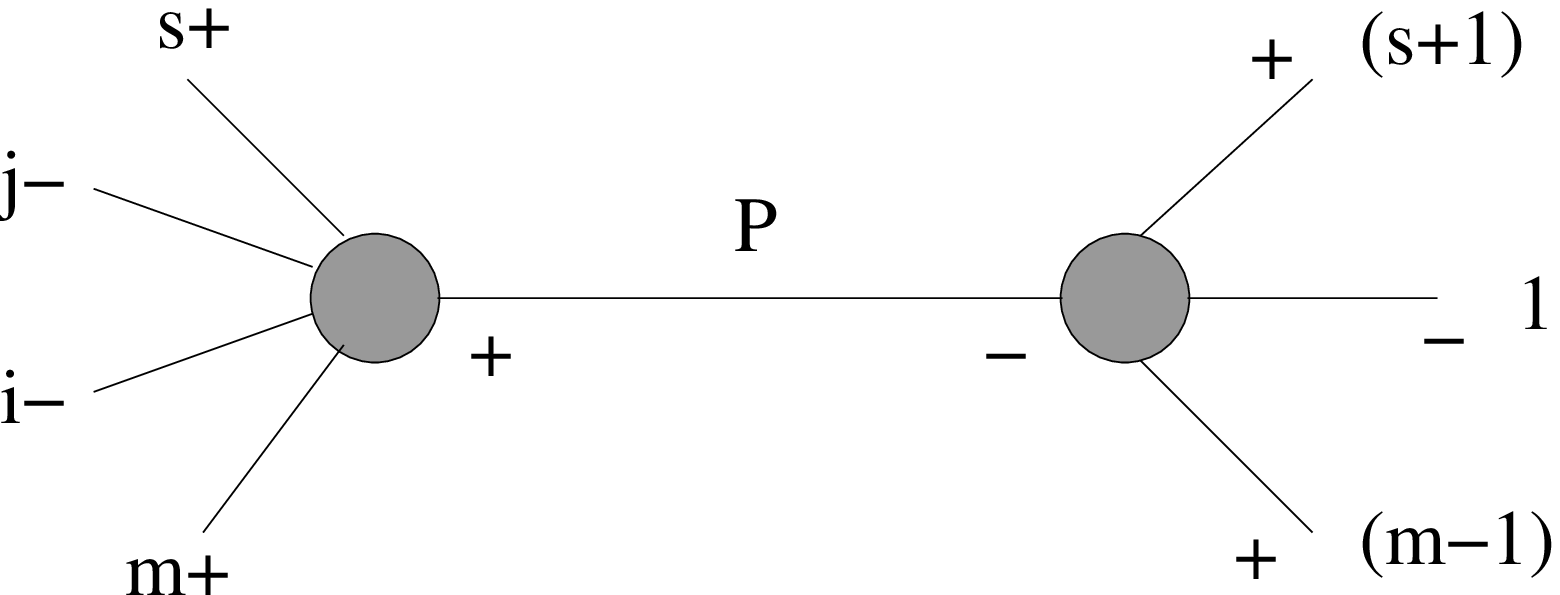}}

In \CachazoKJ, these diagrams were shown to be computed from a
twistor space calculation where gluons are separated into two
groups, with one negative-helicity gluon in one group and two in the other.
An example is shown in Figure 1.
Now, each group is localized on a line and the two lines
are connected by a propagator, as shown in Figure 2. The two lines
do not have to intersect.

\ifig\convi{Twistor diagram that corresponds to the MHV diagram
shown in \convi.} {\epsfxsize=0.55\hsize\epsfbox{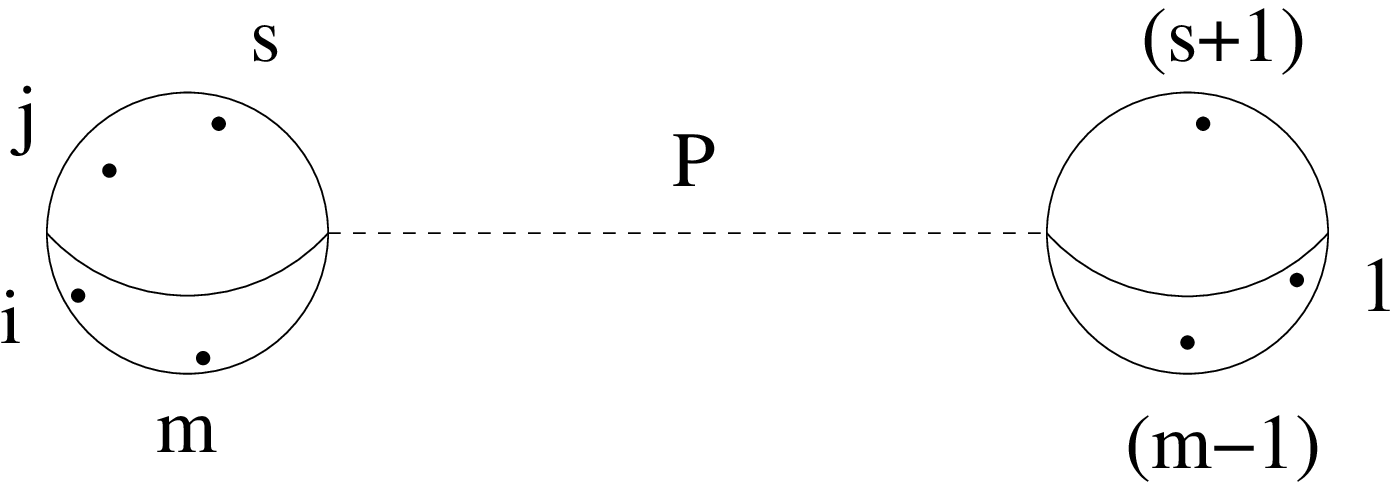}}

From the twistor construction it is clear that all gluons in each group are collinear.

As mentioned above the two lines do not necessarily intersect. The
reason is that the Feynman propagator we use is not well defined
unless an $i\epsilon$ prescription is chosen. The conclusion we
want to get does not depend on the choice so let us just take as
our definition $1/(P^2+i\epsilon)$, where $P=p_m+p_{m+1}+\ldots
+p_s$; see Figure 2. This can be written as the principal value
 plus a delta
function with support at $P^2=0$. Now, the Fourier transform of
the principal part of $1/(P^2+i\epsilon)$ into coordinate space is
a delta function localized at points where $(x-y)^2=0$, where $x$
is the spacetime position of one MHV vertex and $y$ the position
of the other. Therefore, the contribution from the principal value
gives diagrams at points in Minkowski space that are on a light
ray. It turns out that a light ray in twistor space corresponds to a
point \penrose. Recalling that the MHV vertices correspond to
lines in twistor space, this implies that the two lines intersect.
Therefore all gluons are coplanar.

Now we have to worry about the delta function localized at
$P^2=0$. In general we do not discuss these terms because we
consider external gluons with generic momenta and $P^2\neq 0$.
However, in our case, $P$ might contain $\ell_1$ or $\ell_2$,
which are integration variables. It turns out that even in this
case $P^2\neq 0$. To see this, consider $s=\ell_1$ and
$s+1=\ell_2$ in the example of Figures 1 and 2. 
Then $P=p_m+p_{m+1}+\ldots + p_{s-1}+
\ell_1$. The measure in the cut integral has delta functions with
support at 
\eqn\moment{\eqalign{\ell_1^2 &=\ell_2^2=0, \cr
\ell_1+\ell_2 &= -(p_{s+2}+\ldots + p_{m-1}+p_{m} +
\ldots +p_{s-1}).}}
The latter is essentially the momentum conservation constraint for the
NMHV amplitude.

Using \moment\ to solve for $\ell_2$ and imposing that
$\ell_2^2=0$, we find that $R^2 = - 2\ell_1\cdot R$, where
$R=(p_{s+2}+\ldots + p_{m-1}+p_{m} + \ldots +p_{s-1})$. On the
other hand, from $P^2 =0$ we get $Q^2 = - 2\ell_1 \cdot Q $, where
$Q= p_{m} + \ldots +p_{s-1}$.

Because we take generic values of the external momenta 
$p_{s+2},\ldots , p_{m-1}$,
the variables $Q$ and $R$ are independent, and the two equations will 
either localize the integral, producing a rational function, or else cannot
 be 
satisified simultaneously.

\listrefs

\end